\documentstyle[preprint, revtex]{aps}
\def\beq{\begin{eqnarray}}
\def\eed{\end{eqnarray}}
\begin{document}

\begin{title}
\begin{center}
Slave Particle Studies of the Electron Momentum Distribution in the
Low Dimensional $t-J$ Model
\end{center}
\end{title}

\author{Shiping Feng\cite{feng},$^{1}$ J. B. Wu,$^{2}$ Z. B. Su,$^{1,2}$
and L. Yu$^{1,2}$}
\begin{instit}
$^{1}$International Centre for Theoretical Physics, P. O. Box 586,
34100 Trieste, Italy \\
$^{2}$Institute of Theoretical Physics, Chinese Academy of Sciences,
Beijing 100080, China \\
\end{instit}

\begin{abstract}

The electron momentum distribution function in the $t-J$ model
is studied in the framework of slave particle approach. Within the
decoupling scheme used in the gauge field and related theories,
we treat formally  phase and amplitude fluctuations as well as
constraints without further approximations. Our result indicates
that the electron Fermi surface observed in  the high-resolution
angle-resolved photoemission and inverse photoemission
experiments cannot be explained within this framework, and
the sum rule for the physical electron is not obeyed.
A correct scaling behavior of the electron momentum distribution
function near $k \sim k_{F}$ and $k \sim 3k_{F}$ in one dimension
can be reproduced by considering the nonlocal string fields
[Z. Y. Weng et al., Phys. Rev. B45, 7850 (1992)], but the overall
momentum  distribution is still not correct, at least
at the mean field level.
\end{abstract}

PACS   numbers: 71.45. -d, 75.10. Jm

\newpage

\section{Introduction}

The $t-J$ model is one of the simplest models containing the essence
of strong correlations and its implications for oxide
superconductivity \cite{pw1,fcz} still remain an outstanding problem.
The $t-J$ model was originally introduced as an effective
Hamiltonian of the Hubbard model in the strong coupling regime, where
the on-site Coulomb repulsion $U$ is very large as compared with the
electron hopping energy $t$, and therefore the electrons become
strongly correlated to avoid the double occupancy. In this case, the
electron's Hilbert space is severely restricted due to this constraint
$\sum_{\sigma}C^{\dagger}_{i\sigma}C_{i\sigma} \leq 1$.  Anderson
\cite{pw1} and later Zhang and Rice \cite{fcz} have argued strongly
that the basic physics of oxide superconductors can be described by
the $t-J$ model.

The normal state properties of oxide superconductors exhibit a
number of  anomalous properties in the sense that they do not
fit in the conventional Fermi liquid theory \cite{hts,yl1}. Some
properties can be interpreted only in terms of a doped Mott
insulator \cite{hts,yl1}. A central question in the theory of
these strongly correlated systems concerns the nature of the
electron Fermi surface(EFS) \cite{yl1}. The high-resolution
angle-resolved photoemission and inverse photoemission
experiments \cite{olson} demonstrate the existence of a large EFS,
with an area consistent with the band structure calculations.
Since the band theory is consistent with the Luttinger theorem
\cite{jml}, this means that the EFS area contains 1 - $\delta$
electron per site,  where $\delta$ is the hole doping
concentration. Although the topology of EFS is in general
agreement with one-electron-band calculations, the Fermi velocity
is quite different. This indicates that electron correlations
renormalize considerably the results obtained in the framework of
a single-particle treatment. It has recently been shown by small
cluster diagonalization that in a two-dimensional (2D) square
lattice the EFS within the $t-J$ model is consistent with Luttinger's
theorem \cite{wsph}. Monte Carlo simulations
for a nearly half-filled 2D Hubbard model also support this result
\cite{djs}. Moreover, the electron  momentum distribution function
of a 2D $t-J$ model was studied \cite{rvcg} by using the
Luttinger-Jastrow-Gutzwiller variational wave function, which
seems to show the existence of EFS and an algebraic singularity
at the Fermi edge. For one-dimensional (1D) large $U$ limit
Hubbard model which is equivalent to the $t-J$ model,  Ogata and
Shiba \cite{mohs} obtained the electron momentum distribution
function by using Lieb-Wu's exact wave function \cite{ehlw}.
Their result also shows the existence of EFS as well as the
singular behavior at $k\sim k_{F}$ and $k\sim 3k_{F}$ in the
momentum distribution function.  Furthermore, Yokoyama and
Ogata \cite{hyo} studied the 1D $t-J$ model by using exact
diagonalization of small systems and found the power-law
singularity appearing at $k_{F}$ in the momentum distribution
function.

So far, strong correlation effects can be properly taken into
account only by numerical methods \cite{wsph,djs,rvcg,mohs,hyo},
such as variational Monte-Carlo  technique \cite{tkl}, exact
cluster diagonalization \cite{edg}, and  various realizations
of quantum Monte-Carlo method \cite{sorella1}.  Apart from
these numerical techniques,  an analytical approach to the $t-J$
model receiving a great deal of attention is the slave particle
theory \cite{yl1,lee1}, where  the electron  operator $C_{i\sigma}$
is presented as
$C_{i\sigma}$=$a^{\dagger}_{i}f_{i\sigma}$ with $a^{\dagger}_{i}$
as the slave boson and $f_{i\sigma}$ as the fermion, or vice versa.
This way the non-holonomic constraint
\begin{equation}
\sum_{\sigma}C^{\dagger}_{i\sigma}C_{i\sigma}\leq 1
\end{equation}
is converted into a holonomic one
\begin{equation}
a^{\dagger}_{i}a_{i} + \sum_{\sigma}f^{\dagger}_{i\sigma}f_{i\sigma}
= 1 ,
\end{equation}
which means a given site cannot be occupied by more than one particle.
A new gauge degree of freedom must be introduced to incorporate
the constraint, which means that the slave particle representation
should be invariant under a local gauge transformation
$a_{i}\rightarrow a_{i}e^{i\theta (r_{i},t)}$,
$f_{i\sigma}\rightarrow f_{i\sigma}e^{i\theta (r_{i},t)}$, and all
physical quantities should be invariant with respect to this
transformation.  We call such slave particle approach as
"conventional". The advantage of this formalism is that the charge
and spin degrees of freedom  of electron may be  separated at the
mean field level, where the elementary charge and spin  excitations
are called holons and spinons, respectively, but spinons and holons
are strongly coupled by the gauge field (phase) fluctuations
\cite{lee2} or other effects beyond the mean field approximation.
The two fluids of spinons and holons represent the same set of electrons,
and they must, on average, flow together. The decoupling of charge and
spin degrees of freedom in the large $U$ limit Hubbard model is
undoubtedly correct in 1D \cite{mohs}, where the charge degrees of
freedom of the ground state are expressed as a Slater determinant
of spinless fermions, while its  spin degrees of freedom are
equivalent to the 1D $S={1\over 2}$ Heisenberg model, so that
charge and spin excitations propagate at different velocties. However,
the situation is still not clear in 2D.

The 1D $t-J$ model (the large $U$ limit Hubbard model) behaves like
Luttinger liquids \cite{mohs,hyo}. The separation of the spin and
charge degrees of freedom is indeed generic to the universality
class of Luttinger liquids \cite{fdm}. Anderson \cite{pw2} has
hypothesized that  the normal state of 2D strongly interacting
systems relevant for the oxide superconductors should be described
in terms of the theory of Luttinger liquids. Thus it is interesting
to understand the global features of the electron momentum
distribution function  and EFS in the $t-J$ model where separation
of the spin and charge degrees of freedom might be expected.
To our knowledge, the global features of the  electron momentum
distribution function and EFS have been obtained only by using numerical
techniqus, and have not been studied systematically by analytical methods.

In this paper, we study analytically the electron momentum
distribution of the $t-J$ model by using the slave particle approach.
In Sec. II, we present a formal study of the electron momentum
distribution function and discuss the sum rule obeyed by it.
Under the decoupling scheme commonly used in the gauge field
\cite{lee2} and related theories, we treat formally the phase and
amplitude fluctuations, constraints and other effects without
further approximations. The results obtained indicate that the EFS
observed in the high-resolution angle-resolved photoemission and
inverse photoemission experiments \cite{olson} cannot be explained
in the conventional slave particle approach. Moreover, the sum rule
of the physical electron is not obeyed, if the corresponding  sum
rules are imposed on above particles. Since the total number of
electrons is independent of interactions, this result is also valid
beyond the decoupling scheme within the slave particle approach.
In Sec. III, we give an example, following Ref. 20, to show that
the scaling behavior of the electron momentum distribution function
near $k\sim k_{F}$ and $k\sim 3k_{F}$ in 1D can be reproduced by
considering the nonlocal string  fields in the mean field
approximation, but there is still no EFS, and the sum rule for the
electron number is still violated. If Luttinger theorem is obeyed,
the Fermi volume is invariant under interactions. Thus our mean
field result seems to indicate that the correct scaling behavior
of the electron momentum distribution does not guarantee the
existence of EFS (in the sense of global distribution) in the
same theoretical framework. Section IV is devoted to a summary
and discussions on related problems.

\section{Formal Study of Electron Momentum Distribution}
\label{sec:fs}

The slave particle theory can be the slave boson or the slave fermion
theories according to statistics assigned to spinons and holons.
The slave boson formulation \cite{pw3} is one of the popular methods
of treating the $t-J$ model. This method, however, does not give a
good energy of the ground state \cite{pw4}. For example, in the
half-filled case, where the $t-J$ model reduces to the
antiferromagnetic Heisenberg model, the lowest energy state obtained
by this method fails to show the expected long-range N\' eel order,
and the energy is considerably higher than numerical estimates
\cite{tkl}. Also, it does not satisfy the Marshall sign rule
\cite{wm}. However, this theory provides a {\sl spinon} Fermi
surface \cite{lee2} even in the mean field approximation.
Alternatively, the slave fermion approch naturally gives an
ordered N\' eel state at half-filling \cite{dyo}, which obeys
the Marshall \cite{wm} sign rule. The ground state energy obtained
in this case is much better than the slave boson case. At the mean
field level, they are quite different although in princple they
should be equivalent to each other. The differences may be
reduced by going beyond the mean field \cite{yl2} approximation.
In the following, we use both approaches to discuss the electron
momentum distribution function and the sum rule obeyed by it,
and a direct comparison of the obtained results is made.

First consider the slave fermion representation in which
the electron operator can be expressed as
$C_{i\sigma}=e^{\dagger}_{i}b_{i\sigma}$, where $e^{\dagger}_{i}$
is the slave fermion, while $b_{i\sigma}$ is the Schwinger boson,
with the constraint $\sum_{\sigma}b^{\dagger}_{i\sigma}b_{i\sigma}
+e^{\dagger}_{i}e_{i} = 1$. In this case, the Lagrangian $L_{sf}$
and the partition function $Z_{sf}$ of the $t-J$ model in the
imaginary time $\tau$ can be written as
\begin{equation}
L_{sf} = \sum_{i\sigma}b^{\dagger}_{i\sigma}\partial _{\tau}b_{i\sigma}+
\sum_{i}e^{\dagger}_{i}\partial _{\tau}e_{i} + H + \sum_{i}\lambda_{i}
(e_{i}^{\dagger}e_{i} + \sum_{\sigma}b^{\dagger}_{i\sigma}
b_{i\sigma} - 1 ) ,
\end{equation}
\begin{equation}
H = -t\sum_{<ij>\sigma}e_{i}e^{\dagger}_{j}b^{\dagger}_{i\sigma}
b_{j\sigma}+{J\over 4}\sum_{<ij>}
b^{\dagger}_{i\alpha}b_{i\beta}b^{\dagger}_{j\gamma}b_{j\delta}
(\sigma _{\alpha \beta}\sigma _{\gamma \delta} - \delta _{\alpha \beta}
\delta_{\gamma \delta}) - \mu \sum_{i}e^{\dagger}_{i}e_{i} ,
\end{equation}
\begin{equation}
Z_{sf} = \int DeDe^{\dagger}DbDb^{\dagger}D\lambda
e^{-\int d\tau L_{sf}(\tau)} ,
\end{equation}
where $\lambda _{i}$ is the Lagrangian multiplier on site $i$, $\mu$
is the chemical potential, and the summation $<ij>$ is  carried over
nearest neighbors. Following the common practice,  the boson operator
$b^{\dagger}_{i\sigma}$ keeps track of the spin, while the fermion
operator $e^{\dagger}_{i}$ keeps track of the charge, i.e.,
they should obey the following  sum rules:
\begin{equation}
\delta = <e^{\dagger}_{i}e_{i}> = {1\over Z_{sf}}\int DeDe^{\dagger}Db
Db^{\dagger}D\lambda e^{\dagger}_{i}e_{i}e^{-\int d\tau L_{sf}(\tau )} ,
\end{equation}
\begin{equation}
1 - \delta =\sum_{\sigma}<b^{\dagger}_{i\sigma}b_{i\sigma}>
= {1\over Z_{sf}}
\sum_{\sigma}\int DeDe^{\dagger}DbDb^{\dagger}D\lambda
b^{\dagger}_{i\sigma}
b_{i\sigma}e^{-\int d\tau L_{sf}(\tau )} ,
\end{equation}
where $\delta$ is the hole doping concentration, and
$< \cdot \cdot \cdot >$ means the thermodynamical average.
We assume that there is no bose condensation of spinons, which means
that the temperature \cite{dyo} of the system is $T = 0^{+}$.
In the gauge field theory \cite{lee2,ioffe1}, one can introduce
the SU(2)  invariant Hubbard - Stratonovich transformation and
decouple the Lagrangian by using the auxiliary fields. In the
present formal study, there is no need to make any transformation
for the Lagrangian (3), and we will treat formally phase
fluctuations (gauge fields) \cite{lee2}, amplitude fluctuations
\cite{yl2}, constraints et al.  For discussing the electron momentum
 distribution function, we define the following Matsubara fermion,
boson, and electron Green's functions
\begin{eqnarray}
g(R_{i}-R_{j}, \tau -\tau ')
= -<T_{\tau}e_{i}(\tau)e^{\dagger}_{j}(\tau ')> ~~~~~~~~~~~~~ \nonumber \\
= -{1\over Z_{sf}}\int DeDe^{\dagger}DbDb^{\dagger}D\lambda e_{i}(\tau)
e^{\dagger}_{j}(\tau ')e^{-\int d\tau L_{sf}(\tau )} ,
\end{eqnarray}
\begin{eqnarray}
D_{\sigma}(R_{i}-R_{j}, \tau -\tau ')
= - <T_{\tau}b_{i\sigma}(\tau)b_{j\sigma}^{\dagger}(\tau ')> ~~~~~~~~~~~~~
\nonumber \\
= -{1\over Z_{sf}}\int DeDe^{\dagger}DbDb^{\dagger}D\lambda
b_{i\sigma}(\tau)
b^{\dagger}_{j\sigma}(\tau ')e^{-\int D\tau L_{sf}(\tau)} ,
\end{eqnarray}
\begin{eqnarray}
G_{\sigma}(R_{i}-R_{j}, \tau -\tau ')
= -<T_{\tau}C_{i\sigma}(\tau)C^{\dagger}_{j\sigma}(\tau ')> ~~~~~~~~~~~~~~~~
\nonumber \\
= -<T_{\tau}e^{\dagger}_{i}(\tau)e_{j}(\tau ')b_{i\sigma}(\tau)
b^{\dagger}_{j\sigma}(\tau ')> ~~~~~~~~~~~~~~~~~~~~~~~~~~~~~ \nonumber \\
= -{1\over Z_{sf}}\int De De^{\dagger}DbDb^{\dagger}D\lambda
e^{\dagger}_{i}
(\tau)e_{j}(\tau ')b_{i\sigma}(\tau)b^{\dagger}_{j\sigma}(\tau ')
e^{-\int d\tau L_{sf}(\tau)} .
\end{eqnarray}
The spinon-holon scattering contained in eq. (10) is a four particle
process, therefore the spinon and holon are strongly coupled.
At the mean field level, the holons and spinons are separated
completely. However, in many theoretical frameworks, such as the
usual gauge theories discussed by many authors \cite{lee2,ioffe1},
where the vertex corrections are ignored, but the RPA  bubbles are
included, the spinons and holons are still strongly coupled by
phase fluctuations (gauge fields) \cite{lee2}, amplitude
fluctuations \cite{yl2} and other effects. Nevertheless, in all
these cases the electron Green's function $G_\sigma(k, iw_{n})$ can
be presented as a convolution of the fermion Green's function
$g(k,iw_{n})$ and boson Green's function $D_{\sigma}(k,iw_{n})$
\begin{equation}
G_{\sigma}(k,iw_{n})={1\over N} \sum_{q}{1 \over \beta}
\sum_{w_{m}}g(q,iw_{m})D_{\sigma}(q+k, iw_{m}+iw_{n}) .
\end{equation}
The coupling of the gauge field to these particles can be strong
and a partial resummation of digrams has been carried out
\cite{lee2}, but the vertex corrections were neglected, being the
essence of the decoupling approximation. This is an important but
the only approximation apart from the sum rules for slave
particles (6)-(7) in our formal study. In what follows, one will
find that many difficulties might appear due to this approximation.
The fermion and boson Green's functions can be expressed as frequency
integrals of fermion and boson spectral functions as
\begin{equation}
g(k,iw_{n}) = \int_{-\infty}^{\infty}{dw\over 2\pi}
{A_{e}(k,w)\over iw_{n} - w} ,
\end{equation}
\begin{equation}
D_{\sigma}(k,iP_{n}) = \int_{-\infty}^{\infty}
{dw\over 2\pi}{A_{b\sigma}(k,w)\over iP_{n} - w} ,
\end{equation}
respectively \cite{gdm}. Substituting eqs. (12) and (13) into eq. (11),
we obtain the electron Green's function by summing over
the Matsubara frequency $iw_{m}$,
\begin{eqnarray}
G_{\sigma}(k,iw_{n}) = ~~~~~~~~~~~~~~~~~~~~~~~~~~~~~~~~~~~~~~~~~~~~~
{}~~~~~~~~~~~~~~~~~~~~~~~~~~~~~~~~~~ \nonumber \\
{1\over N}\sum _{q}\int_{-\infty}^{\infty}{dw' \over 2\pi}
\int_{-\infty}^{\infty} {dw'' \over 2\pi} A_{e}(q,w')
A_{b\sigma}(q+k,w''){n_{F}(w')+n_{B}(w'')\over iw_{n}+w'-w''} ,
\end{eqnarray}
where $n_{F}(w')$ and $n_{B}(w'')$ are the Fermi and Bose distribution
functions, respectively. The spectral functions $A_{e}(q,w')$ and
$A_{b\sigma}(k,w'')$ obey the sum rules coming from the commutation
relations,
\begin{equation}
\int_{-\infty}^{\infty} {dw \over 2\pi} A_{e}(q,w) =1 ,
\end{equation}
\begin{equation}
\sum_{\sigma} \int_{-\infty}^{\infty} {dw\over 2\pi} A_{b\sigma}(k,w)
=2 .
\end{equation}
The electron's Hilbert space has been severely restricted,  but the
fermion and boson themselves are not restricted. By an analytic
continuation $iw_{n}\rightarrow w + i\eta$ in the electron Green's
function(14), the electron
spectral function $A_{c\sigma}(k,w) = -2ImG_{c\sigma}(k,w)$
can be obtained as
\begin{eqnarray}
A_{c\sigma}(k,w) = ~~~~~~~~~~~~~~~~~~~~~~~~~~~~~~~~~~~~~~~~~~~~~~~~~~~
{}~~~~~~~~~~~~~~~~~~~~~~~ \nonumber \\
{1\over N} \sum_{q}\int_{-\infty}^{\infty}{dw'\over 2\pi} A_{e}(q,
w')A_{b\sigma}(q+k,w+w')[n_{F}(w')+n_{B}(w+w')] .
\end{eqnarray}
Therefore, the electron spectral function $A_{c\sigma}(k,w)$
obeys the sum rule
\begin{equation}
\sum_{\sigma}\int_{-\infty}^{\infty}{dw\over 2\pi}
A_{c\sigma}(k,w)=1+\delta ,
\end{equation}
which is less than 2 since an amount of (1 - $\delta$) of the doubly
occupied Hilbert space is pushed to infinity as $U \rightarrow \infty$
in deriving the $t-J$ model. Thus the spectral function
$A_{c\sigma}(k,w)$ only describes the lower Hubbard band.
In deriving eq. (18), we have used the identities
\begin{equation}
n_{b\sigma}(k)=\int_{-\infty}^{\infty}{dw\over 2\pi}
n_{B}(w)A_{b\sigma}(k,w) ,
\end{equation}
\begin{equation}
n_{e}(k)=\int_{-\infty}^{\infty}{dw\over 2\pi}n_{F}(w)A_{e}(k,w) .
\end{equation}
This is because $n_{B}(w)A_{b\sigma}(k,w)$ and
$n_{F}(w)A_{e}(k,w)$ can be interpreted as the
probability functions of state $k$ with energy $w$ for boson
and fermion, respectevely. A similar interpretion is also valid for
the electron spectral function $A_{c\sigma}(k,w)$. Thus the number
of electrons in state $k$ is obtained by summing over all energies
$w$, weighted by the electron spectral function
\begin{equation}
n_{c}(k)=\sum_{\sigma}\int_{-\infty}^{\infty}{dw\over 2\pi}
n_{F}(w)A_{c\sigma}
(k,w)=1-\delta-{1\over N}\sum_{q\sigma}n_{b\sigma}(k+q)n_{e}(q) .
\end{equation}
In eq.(21), the first term of the righthand side 1 - $\delta$ is
independent of $k$, and therfore it is true for all $k$ states
of the entire Brillouin zone. The value of the second term of
the righthand side is of the order of $\delta$, and hence
it is not enough to restore the EFS, i.e., the distribution outside
the  should-be  EFS is still of the order 1. Fig. 1 shows the mean
field electron momentum distribution $n_{c}(k)$ for doping
$\delta =0.125$ in 1D. Beyond the mean field approximation,  but
still within the decoupling scheme (11), there are no important
corrections for the global features of the electron momentum
distribution and EFS, but the Fermi velocity will be modified.
This is because the essential global features of the electron
momentum distribution are dominated by the first term of the
righthand side,  $1-\delta$ in eq. (21), and the second term of the
righthand side in eq. (21), which is of the order of $\delta$,
is not enough to cancel out $1-\delta$ at each $k$ outside the
$k_{F}$ state, i.e., $k_{F}<k<\pi$, for $k>0$, and $-\pi<k<-k_{F}$,
for $k<0$. Since $n_{b\sigma}(k+q)\ge 0$ and $n_{e}(q)\ge 0$, and a
minus sign appears between the first and the second terms of the
righthand side in eq. (21), then it is impossible to shift the weight
$1-\delta$ in state $k$ outside the Fermi points into state $k'$
inside the Fermi points, i.e., $0<k'<k_{F}$, for $k'>0$, and
$-k_{F}<k'<0$, for $k'<0$. The above discussions are also true for
the 2D case. Therefore, there is no EFS in the standard sense
within the decoupling scheme (11) in the conventional slave fermion
approach. If $\delta$ holes are introduced into the half-filled system,
one might expect that the total electron number per site would be
1 - $\delta$. However, a surprising result is
\begin{equation}
{1\over N}\sum_{k}n_{c}(k)=(1-\delta)^{2} ,
\end{equation}
which is not the expected value, and violates the sum rule of the
electron number.

Alternatively, in the slave boson representation, the electron
operator can be expressed as $C_{i\sigma}=f_{i\sigma}b_{i}
^{\dagger}$,  where $f_{i\sigma}$ is fermion and $b^{\dagger}_{i}$
is slave boson,  with the contraint $\sum_{\sigma}
f_{i\sigma}^{\dagger}f_{i\sigma}+
b_{i}^{\dagger}b_{i}=1$. In this case, the Lagrangian $L_{sb}$
and the partition function $Z_{sb}$ of the $t-J$ model may be
written as
\begin{equation}
L_{sb} = \sum_{i\sigma}f_{i\sigma}^{\dagger}\partial_{\tau} f_{i\sigma}
+ \sum_{i}b_{i}^{\dagger}\partial_{\tau} b_{i} + H +\sum_{i}\lambda_{i}
(b^{\dagger}_{i}b_{i} + \sum_{\sigma}f^{\dagger}_{i\sigma}
f_{i\sigma} - 1) ,
\end{equation}
\begin{eqnarray}
H = -t\sum_{<ij>\sigma}b_{i}b^{\dagger}_{j}f^{\dagger}_{i\sigma}
f_{j\sigma}
- \mu \sum_{i\sigma}f^{\dagger}_{i\sigma}f_{i\sigma} \nonumber \\
+ {J\over 4}\sum_{<ij>}f^{\dagger}_{i\alpha}f_{i\beta}
f^{\dagger}_{j\gamma}
f_{j\delta}(\sigma_{\alpha \beta}\sigma_{\gamma \delta} -
\delta_{\alpha \beta}\delta_{\gamma \delta}) ,
\end{eqnarray}
\begin{equation}
Z_{sb} = \int DbDb^{\dagger}DfDf^{\dagger}D\lambda
e^{-\int d\tau L_{sb}(\tau)} ,
\end{equation}
where the fermion operator $f_{i\sigma}^{\dagger}$ keeps
track of the spin, while the boson operator $b_{i}^{\dagger}$
keeps track of the charge, i.e.,
\begin{equation}
\delta = <b^{\dagger}_{i}b_{i}> ={1\over Z_{sb}}\int DbDb^{\dagger}
DfDf^{\dagger}
D\lambda b_{i}^{\dagger}b_{i}e^{-\int d\tau L_{sb}(\tau)} ,
\end{equation}
\begin{equation}
1 - \delta =\sum_{\sigma}<f^{\dagger}_{i\sigma}f_{i\sigma}>
={1\over Z_{sb}}
\sum_{\sigma}\int DbDb^{\dagger}DfDf^{\dagger}D\lambda
f^{\dagger}_{i\sigma}
f_{i\sigma}e^{-\int d\tau L_{sb}(\tau)} .
\end{equation}
We assume that there is no bose condensation of holons. Strictly
speaking, this is true above the Bose condensation temperature.
However, as shown by Lee and Nagaosa \cite{lee2}, the
inelastic scattering of bosons by the gauge field suppresses
significantly the Bose condensation temperature. After some formal
calculations which are similar to the slave fermion  case, we obtain
\begin{equation}
\sum_{\sigma}\int_{-\infty}^{\infty}{dw\over 2\pi}
A_{c\sigma}(k,w) =1 +\delta ,
\end{equation}
\begin{equation}
n_{c}(k)=1-\delta +{1\over N}\sum_{q\sigma}n_{f\sigma}(k+q)n_{b}(q) ,
\end{equation}
\begin{equation}
{1\over N}\sum_{k}n_{c}(k)=1 - \delta^{2} .
\end{equation}
In comparison with eq. (21), the second term of the righthand side
of eq. (29) changes sign as an essential difference of the electron
momentum distribution function between the slave boson and slave
fermion representations. However, the value of the second term of
the righthand side of eq. (29) is also of the order of $\delta$,
and it is also not enough to restore the EFS. Fig. 2 shows the mean
field electron momentum distribution
$n_{c}(k)$ for doping $\delta =0.125$ in 1D. Beyond the mean field
approximation, but still within the decoupling scheme, there are no
important corrections for the global features of the electron
momentum distribution. The reason is  almost the same as in the
conventional slave fermion approach. The essential global features
of the electron momentum distribution are dominated by the
first term of the righthand side $1-\delta$ in eq. (29). Since
$n_{f\sigma}(k+q)\ge 0$ and $n_{b}(q)\ge 0$, and a plus sign appears
between the first and the second terms of the righthand side in
eq. (29), it is impossible to remove those $1-\delta$ states beyond
the Fermi points. The best situation is that an amount of order of
$\delta$ is added into each $k'$ state within the Fermi points.
The above discussion is also valid for the 2D case. Therefore, there
is no EFS  within the decoupling scheme in the conventional slave
boson approach. In this case, the sum rule of the electron number
is also violated as in the slave fermion case. The difference between
these two approaches is that the electron number is more than the
expected value in the slave boson representation, but less than it in
the slave fermion representation.

These results indicate that there is no  real EFS for the electron
momentum distribution function $n_{c}(k)$ near $k \sim k_{F}$ within
the decoupling scheme in the  conventional slave particle approach.
The sum rule of the physical electron number is violated. The scaling
behavior of the electron momentum distribution near $k_{F}$ is also
not correctly described by this scheme. The theory can be applied to
both 1D and 2D systems. It was proposed \cite{lee2,ioffe1} that the
low energy physics of the $t-J$ model can be described by a theory of
fermions and bosons coupled by a gauge field.  Our results also
indicate that the gauge fields (phase fluctuations) discussed by many
authors \cite{lee2,ioffe1,ioffe2} are not strong enough to restore
the EFS under the decoupling scheme (11). This is because the effects
of phase fluctuations in the slave particle approach are essentially
local in the momentum space. On the other hand, it has been shown from
experiments \cite{olson} that oxide superconducting materials exhibit
an EFS, and the EFS obeys Luttinger theorem \cite{jml}, which means
the Fermi volume is invariant under interaction. In this sense the
EFS of strong correlated systems should also be described by an
adequate theory even in the mean field approximation. In fact, the
rearrangements of spin configuations in the electron hopping process
\cite{mohs}, which are nonlocal effects and beyond the conventional
slave particle approach, play an essential role to restore EFS for a
system of decoupled charge and spin degrees of freedom. Thus the
electron is not a composite of holon and spinon only, and it should
include other fields which describe the nonlocal effects. In the next
Section, we will see that the scaling behavior of $n_{c}(k)$ near
$k\sim k_{F}$ and $k\sim 3k_{F}$ can be obtained in 1D by considering
some nonlocal effects. Before going to the next Section, we would
like to emphasize that the "Fermi Surface" obtained previously
\cite{wang} from the slave boson theory is a {\sl spinon} Fermi
surface, not the real EFS. In order to interpret the high-resolution
angle-resolved photoemission and inverse photoemission experiments
\cite{olson}, the {\sl spinon} Fermi surface is not enough because
experiments have shown the EFS for real electrons.

\section{The Scaling Behavior of the Electron Momentum Distribution \\
Function and the Effects of Nonlocal String Field in 1D}

Interacting 1D electron systems generally behave like Luttinger liquids
\cite{fdm} in which the correlation functions have power-law decay with
exponents which depend on the interaction strength. For the 1D Hubbard
model, an exact solution was explicitly obtained by Lieb and Wu
\cite{ehlw}. In the limit of $J\rightarrow 0$, the $t-J$ model is
equivalent to the large $U$ limit Hubbard model \cite{pw1,fcz,mohs}.
Thus the 1D $t-J$ model provides a good test for various approaches.
The asymptotic forms of some correlation functions as well as the
single electron Green's function have been obtained by many authors
using different approximations \cite{weng,sorella2}. In particular,
Ogata and Shiba \cite{mohs} obtained the electron momentum
distribution function $n_{c}(k)$ by using the Bethe-Ansatz Lieb-Wu
\cite{ehlw} wave function, and their results show the presence of
EFS and the correct scaling behavior of $n_{c}(k)$
at $k\sim k_{F}$ and $k\sim 3k_{F}$.  It is well established that the
Landau Fermi liquid theory breaks down in 1D, namely,
(1) there is no finite
jump of the momentum distribution at the Fermi surface; (2) there is
no quasi-particle propagation and (3) the spin and charge are
separated. On the other hand, there is still a well defined Fermi
surface at $k_{F}$, as one would expect from the Luttinger theorem.
It is remarkable that the exact solution of the 1D Hubbard model
demonstrates explicitly these two aspects at the same time. In this
sense it is important to check whether both aspects remain in
any appropriate approximate treatment of the 1D model.
To our knowledge, such
global features of the electron momentum distribution function even
in 1D were obtained only by using numerical techniqes. In this
Section, we try to study analytically this problem. A CP$^{1}$
boson-soliton approach including the effects of the nonlocal
string field to study the large $U$ Hubbard model was recently
developed by Weng et al. \cite{weng}. They have shown that the
electron is a composite particle of holon, spinon, together with
a nonlocal string field. We will draw heavily on their results,
but we try to make the presentation self-contained. The correct
scaling form of the electron momentum distribution function can be
obtained even in the mean field approximation if one considers the
nonlocal string field.

For convenience, we begin with the $t-J$ model, but consider the
limit $J \rightarrow 0^{+}$ which  is a fixed point different from
$J=0$. In this case, the $t-J$ Hamiltonian may be written as
\begin{equation}
H = -t\sum_{<ij>\sigma} C^{\dagger}_{i\sigma}C_{j\sigma} + h. c.
\end{equation}
As in the CP$^{1}$ boson-soliton approach \cite{weng}, the electron
operator $C_{i\sigma}$ can be expressed in the slave  fermion
representation including the effects of nonlocal string fields as
\begin{equation}
C_{i\sigma} =e^{\dagger}_{i}e^{-i{\pi \over 2}\sum_{l<i}
n_{l}}b_{i\sigma}G_{i} ,
\end{equation}
with the constraint $\sum_{\sigma}b^{\dagger}_{i\sigma}b_{i\sigma} +
e^{\dagger}_{i}e_{i} = 1$.
Here $n_{l}=e^{\dagger}_{l}e_{l}$, the fermion operator
$e^{\dagger}_{i}$ keeps track of the charge, while the boson operator
$b^{\dagger}_{i\sigma}$ keeps track of the spin.
$e^{-i{\pi \over 2}\sum_{l<i}n_{l}}$ is the string field,  which
describes the effects of rearrangements of spin configurations
from $-\infty$ to site $i$ when one electron was removed or added
at site $i$. $G_{i}$ is a projection oprator, which ensures that
$b_{i\sigma}$ annihilates the spin $\sigma$ and is a nonlocal phase
shift $G_{i}=e^{i{\pi \over 2}(N - X_{i}-\sum_{l>i}n_{l})}(X_{i}$
is the lattice site), which describes the effects of rearrangements
of the spin configurations from  site $i$ to $+\infty$ when one
electron was removed or added at site $i$.  As shown by Ogata and
Shiba \cite{mohs}, the "{\sl squeezing}" effect, i.e., the
rearrangement of spins when holes are squeezed out, is crucial in
recovering the Fermi surface and getting the correct exponents for
the correlation functions. The nonlocal "string" proposed in
\cite{weng}, will partly take into account this effect.
In fact the motion of holons is also affected by the
rearrangements of the spin configurations \cite{yl3}. We have
neglected this second effect since the kinetic energy $t$ is much
larger than the magnetic energy $J$. One can check easily that
the anticommunication relations are the same
as for the conventional slave fermion approach. In this case,
the Hamiltonian  with  contraints can be written as
\begin{eqnarray}
H = -t\sum_{j\sigma}[ib^{\dagger}_{j\sigma}b_{j+1\sigma}
e^{\dagger}_{j+1}e_{j}
+ (-i)b^{\dagger}_{j+1\sigma}b_{j\sigma}e^{\dagger}_{j}e_{j+1}]
\nonumber \\
-\mu \sum_{j}e^{\dagger}_{j}e_{j} +\sum_{j}\lambda_{j}
(e^{\dagger}_{j}
e_{j}+\sum_{\sigma}b^{\dagger}_{j\sigma}b_{j\sigma} -1) ,
\end{eqnarray}
where $\mu$ is the chemical potential, and $\lambda_{j}$ is the
Lagrangian multiplier on site $j$. The factor $i$ in the
Hamiltonian is very important, which will shift the holon energy
spectrum and will give rise to the  correct scaling form of the
electron momentum distribution function. Thus we define the new
holon operator as
\begin{equation}
e_{j}=e^{i{\pi \over 2}X_{j}}h_{j} ,
\end{equation}
and the Hamiltonian may be rewritten as
\begin{eqnarray}
H=-t\sum_{i\sigma}[b^{\dagger}_{i\sigma}b_{i+1\sigma}
h^{\dagger}_{i+1}
h_{i}+h.c ] -\mu \sum_{i}h^{\dagger}_{i}h_{i} \nonumber \\
+\sum_{i} \lambda_{i}(h^{\dagger}_{i}h_{i}+\sum_{\sigma}
b^{\dagger}_{i\sigma}b_{i\sigma} -1) .
\end{eqnarray}

Following Weng et al. \cite{weng}, we can obtain the asymptotic
singular behavior of the electron momentum distribution by a similar
calculation, and the result is in agreement with theirs. In doing so,
the major approximation is to drop the $G_{i}$ factor in eq. (32),
while calculating the asymptotic single electron Green's function.
As pointed out by Weng et al. \cite{weng}, this factor will contribute
an additional power-law decay in the asymptotic single electron
Green's function,  and one may negelect it for simplicity if only
interested in the leading contribution.  The main effect due to
the nonlocal string is already present. In the MFA, our situation
is very similar. Eq. (32) holds exactly
in the N\' eel limit of the spin configuration, and therefore
the quantum many-body effects are
neglected. This does not affect the leading long-wavelength behavior.
In this paper, however, we are mainly interested in the
global features of the electron momentum distribution. If Luttinger
theorem \cite{jml} is obeyed, the Fermi volume is invariant under
interaction and a strongly interacting system should also
show a large EFS even in the mean field approximation. Thus
a mean field treatment is a useful test for the present approach.

The mean field approximation to the Hamiltoian(35) amounts to treating
$\lambda_{i}$ as a constant, independent of position and to decoupling
the spinon-holon
interaction in a Hartree-like approximation by introducing the order
paramerters
\begin{equation}
\chi = \sum_{\sigma}<b^{A \dagger}_{i\sigma}b^{B}_{i+1\sigma}> ,
\end{equation}
\begin{equation}
\phi = <h^{A\dagger}_{i}h^{B}_{i+1}> ,
\end{equation}
where we have considered two sublattices A and B with
$i\in A, i+1\in B$. The self-consistent equations about
$\lambda,  \chi,  \phi$ , and $\mu$ can be obtained by
minimizing the free energy.

Under the same approximation as the one used to obtain the
asymptotic single electron Green's function(i.e., $G_{i}$
factor in eq. (32) has been dropped), the single particle
electron Green's function $G_{c\sigma}(k,iw_{n})$ in the
present  mean field approxmation can be obtained as
\begin{equation}
G_{c\sigma}(k,iw_{n})={1\over 2N}\sum_{q\pm}{n_{B}(w_{q})+
n_{F}(\varepsilon_{q-k\pm {\pi\over 2}(1+\delta)})
\over iw_{n}+\varepsilon_{q-k\pm {\pi\over 2}(1+\delta)} - w_{q}} ,
\end{equation}
where
\begin{equation}
w_{k}= \lambda - t\phi \gamma_{k},~~~~~ \varepsilon_{k} =
\lambda -\mu -t\chi \gamma_{k},~~~~~ \gamma_{k} = 2cos(k) .
\end{equation}
Therefore, one can get the electron spectral and momentum
distribution functions
\begin{equation}
A_{c\sigma}(k,w)={\pi\over N}\sum_{q\pm}[n_{B}(w_{q})+n_{F}
(\varepsilon_{q-k \pm {\pi\over 2}(1+\delta)})]
\delta(w+\varepsilon_{q-k\pm {\pi\over 2}(1+\delta)} -w_{q}) ,
\end{equation}
\begin{equation}
A_{c}(k)=\sum_{\sigma}\int_{-\infty}^{\infty}{dw\over 2\pi}
A_{c\sigma}(k,w)=1 + \delta ,
\end{equation}
\begin{equation}
n_{c}(k)=1-\delta -{1\over N}\sum_{q\pm}n_{B}(w_{q})
n_{F}(\varepsilon_{q-k \pm {\pi\over 2}(1+\delta)}) ,
\end{equation}
\begin{equation}
{1\over N}\sum_{k}n_{c}(k)=(1- \delta)^{2} .
\end{equation}
In comparison with eq. (21), we find the holon energy spectrum has
been shifted by ${\pi \over 2}(1+\delta )$ due to the presence of
nonlocal string fields.  Fig. 3 shows the electron momentum
distribution $n_{c}(k)$ in the mean field approximation,
including the nonlocal string fields for doping $\delta =0.125$ in 1D.
The singular behavior of the electron momentum distribution
function at $k_{F}$ and $3k_{F}$  is in qualitative
agreement with the numerical results \cite{mohs},
but the sum rule for the total electron number is still
violated and EFS still does not exist in the present framework
under the above approximations. Beyond these  approximations,
the situation is not clear yet. If Luttinger theorem \cite{jml}
is obeyed, then there are no important corrections to the global
features of the electron momentum distribution function and
no EFS beyond the mean field approximation in the same theoretical
framework. Since the rearrangements of quantum spin configurations
in the hopping process play an essential role to restore the
EFS \cite{mohs}, then eq. (32) which holds  exactly \cite{weng}
in the N\' eel limit of the spin configuration, is perhaps
not enough to describe all of these quantum spin
configurations. Thus our feeling is that in this theoretical framework,
the correct singular behavior of electron momentum distribution may be
obtained, but it does not guarantee the existence of EFS.
We also note that only the asymptotic forms were discussed in
the previous works by many authors \cite{sorella2},  while the global
features of the momentum distribution function have not been
considered by their theoretical methods.

\section{Summary and Discussions}

We have discussed the electron momentum distribution function and
the sum rule of the electron number in the $t-J$ model by using the
slave particle approach. Under the decoupling scheme (11) used in
the gauge field \cite{lee2} and related theories, we have formally
proved that EFS can not be restored and the sum rule of electron
number is violated in the framework of conventional slave particle
approach. For 1D $t-J$ model, the correct singularity forms can be
reproduced by considering the nonlocal string fields in the special
mean field approximation, but there is  still no EFS
and the sum rule is still not satisfied.

In the present slave particle approach to study $t-J$ model, we
have pushed out the upper Hubbard band. This means we have
neglected the doubly occupied sites in the original slave particle
approach \cite{pw3}: $C_{i\sigma}=a^{\dagger}_{i}
f_{i\sigma} + \sigma d_{i}f^{\dagger}_{i-\sigma}$, with a constraint
$a^{\dagger}_{i}a_{i} + \sum_{\sigma}f^{\dagger}_{i\sigma}f_{i\sigma}
+ d^{\dagger}_{i}d_{i} =1$, where $a^{\dagger}_{i}$ is the slave
boson, $f_{i\sigma}$ is fermion, and $d_{i}$ is boson which
describes the doubly occupied sites, or vice versa. In fact in the
original slave particle approach the electron spectral function
obeys the sum rule of conventional electron, i.e.,
$\sum_{\sigma}\int^{\infty}_{-\infty}
{dw\over 2\pi}A_{c\sigma}(k,w) = 2$, but the sum rule of the
electron number is still violated. We also find that the sum rule of
electron number is not violated in the $CP^{1}$ representation of
electrons, which probably means it is a better choice to study
the $t-J$ model.

The decoupling of charge and spin degrees of freedom of electron is
undoubtedly correct in 1D $t-J$ model in the $J\rightarrow 0^{+}$
limit \cite{mohs}, and 1D $t-J$ model behaves as Luttinger liquids
\cite{mohs,hyo}. Anderson \cite{pw2} has hypothesized that the
normal state of 2D strongly interacting systems relevant for
the oxide supercondutors might show  generalized Luttinger liquid
behavior, and the characteristic separation of charge and spin
excitations might be responsible for the experimentally observed
temperature dependences of the resistivity \cite{fiory} and Hall
effect \cite{ong}. If these properties can be described in the
framework of slave particle approach, then we must be out of the
difficulties mentioned in Sec. II and Sec. III. A possible way to
restore the EFS is to try to include the vertex corrections
beyond the decoupling scheme (11). As stated earlier, the coupling
to the gauge field can be strong, but so long as the vertex
corrections are  neglected, i.e., the decoupling scheme is
adopted, the difficulty will remain.  On the other hand, it is
difficulty to introduce vertex corrections in the present
form of the gauge field theory, because the infrared
divergence has not yet been properly handled.
Another possibility to avoid this
difficulty is that one should make a new interpretation of the
constraint (2) and the physical meaning of spinons and holons.
In fact the constraint (2) is an operator identity, and one
replaces the constraint (2) by eqs. (6) and (7) in the slave
fermion approach, or eqs. (26) and (27) in the slave boson case.
It is not clear that this is a correct way of imposing the
constraint, i.e., the charge is represented by a fermion, while
the spin is represented by a boson in the slave fermion
approach and vice versa in the slave boson version. The
crucial point is to implement the local constraint rigorously
\cite{feng6}. Finally, according to the implications
of numerical solutions \cite{mohs},  the electron is perhaps not a
composite of holon and spinon only, but it rather should be a
composite of holon, spinon together with something else. This
is because the Bethe-ansatz Lieb-Wu's exact wave function \cite{mohs}
for 1D Hubbard model at large $U$ limit may be written as
\begin{equation}
\Psi (x_{1},\cdot \cdot \cdot ,x_{N}) = (-1)^{Q}
det[exp(ik_{i}x_{Q_{i}})] \Phi(y_{1},\cdot \cdot \cdot ,y_{M}) ,
\end{equation}
where the determinant depends only on the coordinates of
particles ($x_{Q_{1}}< \cdot \cdot \cdot <x_{Q_{N}}$) but not on
their spins. Thus it is the same as the Slater determinant of
spinless fermions with momenta $k_{j}$'s. The spin wave function
$\phi (y_{1},\cdot \cdot \cdot ,y_{M})$ is the same as the
Bethe's exact solution of 1D Heisenberg spin system. In the
second quantization representation, a spinless fermion operator
may be responsible for the Slater determinant of spinless fermions,
but a simple fermion or boson operator may be not enough to describe
the spin wave function. Perhaps it should be a fermion or boson
operator together with something else, such as the Jordan-Wigner's
form \cite{jordan} in 1D, or a map of the three spin operators of
the SU(2) algebra onto a couple of canonical boson operators, like
the Holstein-Primakoff form \cite{hp}, are responsible for the spin
wave function. We believe that a successful theoretical framework
must include the essential ingredients to give a qualitatively
correct description of the global features of EFS even in the mean
field approximation.

To summarize, we are facing a dilemma: if we would like to split
one electron into two particles, one keeping track of the charge,
the other keeping track of the spin, and impose corresponding sum
rules on them [eqs. (6)-(7) or eqs. (26)-(27)] and then use the
decoupling scheme for their expectation values, the resulting
electron distribution does not satisfy the sum rule and does not
show an EFS. In this sense, we have proved in this paper a No-Go
theorem. The alternative would be to give up the attractively simple
interpretation of spinons and holons and to look for more
complicated charge and spin collective excitations.

Finally we note that the  present slave particle approach is
different from those first proposed by Barnes \cite{seb} and
rediscovered and extended by Coleman \cite{pc}, Read and Newns
\cite{read}, and Kotliar and Ruckenstein \cite{li} in  their
works on the mixed valence problem and the heavy fermion systems.
In their formulation, the spin and charge degrees of freedom are
not decoupled, where the auxiliary bosons keep track only of the
environment by measuring the occupation  numbers in each of the
possible states for electron hopping. Their theory describes
the properties of a Fermi liquid, and is another story.

\acknowledgements

The authors would like to thank Dr. Z. Y. Weng  for helpful
discussions. S. P. Feng and Z. B. Su would like to thank the
ICTP for the hospitality.

\figure{The mean field electron momentum distribution $n_{c}(k)$ for
doping $\delta =0.125$ in 1D in the conventional slave fermion
approach. The integrated area of the electron momentum distribution is
$1 -\delta -\delta(1 - \delta) = (1 - \delta )^{2}$. }
\figure{The mean field electron momentum distribution $n_{c}(k)$ for
doping $\delta =0.125$ in 1D in the conventional slave boson approach.
The integrated area of the electron momentum distribution is
$1 -\delta +\delta(1 - \delta) = 1 - \delta^{2}$ }
\figure{The mean field  electron momentum distribution $n_{c}(k)$ for
doping $\delta =0.125$ in 1D in the slave fermion approach including the
nonlocal string fields. The integrated area of the electron
momentum distribution is $(1 - \delta )^{2}$.  Note that, for
$\delta =0.125$, $3k_{F}$ becomes larger than $\pi$ and the
singularity appears at $3k_{F}$-$2\pi$, while -$3k_{F}$ is less
than -$\pi$ and the singularity appears at -$3k_{F}$+$2\pi$.
The values of $k_{F}$ and -$3k_{F}$+$2\pi$ are indicated by arrows. }

\end{document}